\documentstyle{lamuphys}
\begin{document}
\title{
{\hspace{9cm}
\rm \small 
HUB-EP-97/80}
\\ \medskip\\
Hadronization in Particle Physics}
\titlerunning{Hadronization in Particle Physics}
\author{D.Ebert\thanks{Invited talk given at the Workshop ``Field 
Theoretical Tools in Polymer and Particle Physics'', University Wuppertal, 
June 17-19, 1997}}
\institute {Institut f\"ur Physik, Humboldt-Universit\"at zu Berlin,\\ 
Invalidenstrasse 110, D-10115, Berlin, Germany}
\maketitle
\begin{abstract}
The method of path integral hadronization is applied to a local 
quark-diquark toy model in order to derive an effective chiral 
meson-baryon Lagrangian. Further generalizations to models including 
both scalar and axial-vector diquarks as well as nonlocal interactions 
are discussed.
\end{abstract}
\section{Introduction}
In the first lecture (referred to as {\em I} in the following) I have 
shown that QCD-motivated effective quark models of the NJL type can be 
reformulated as effective theories given in terms of composite bosonic 
objects, mesons. An analogous treatment of baryons as relativistic 
bound systems leads us in the case $N_{\rm c}=3$ to the concept of diquarks. 
Diquarks as effective degrees of freedom have been introduced both 
in two-dimensional QCD [1] and in four-dimensional QCD-type models [2]. 
By applying path integral methods to an NJL-type model with 2-body 
$\left(q\tilde q\right)$ and $\left(qq\right)$ forces it was, in 
particular, possible to derive Faddeev equations determining the 
spectrum of composite baryons [3-5]. It is a further challenge to derive 
the well known effective chiral meson-baryon Lagrangians [6,7], which 
reproduce low energy characteristica of hadron physics directly by the 
path integral hadronization method. 

In this second lecture I will demonstrate how this can be done by 
choosing a simple quark-diquark toy model containing a local 
interaction of elementary scalar diquarks with quarks. Finally, possible 
generalizations to models containing scalar and axial-vector diquarks 
taken as elementary fields [8] or composite fields and including 
nonlocal interactions mediated by quark exchange [3-5,9] are discussed. 

\section{Hadronization of a Quark-Diquark Toy Model}
Let us consider a chiral-invariant Lagrangian containing the 
semi-bosonized meson-quark Lagrangian ${\cal L}_{{\rm NJL}}^{\chi 
{\rm M}}$ described in lecture {\em I} (see (22)) supplemented 
by a diquark and an interaction term  

\begin{equation}
{\cal L}^{\chi {\rm MD}}={\cal L}_{{\rm NJL}}^{\chi {\rm M}}+
D^\dag\left(-\Box-M_D^2\right)D+\tilde G\left(\bar\chi D^\dag\right)
\left(D\chi\right).
\end{equation}
Here $D(x)$ is the field of an (elementary) scalar isoscalar diquark 
of mass $M_D$, and $\tilde G$ is a coupling constant.

Analogously to the introduction of collective meson fields into the 
NJL model (Cf. (4) of {\em I}) let us now introduce collective 
baryon (nucleon) fields $B$ by using the identity

\begin{equation}
\E^{\I\int \D^4x\tilde G\left(\bar\chi D^\dag\right)\left(D\chi\right)}
={\cal N}'\int {\cal D}B{\cal D}\bar B\E^{\I\int \D^4x\left(-\frac{1}
{\tilde G}\bar B B-\bar\chi D^\dag B-\bar B D\chi\right)}.
\end{equation}
The ``hadronization'' of the generating functional of the Lagrangian (1) 
will now be performed step by step by integrating over the microscopic 
quark and diquark fields. We obtain    

$$
{\cal Z}={\cal N}_1\int {\cal D}\mu\left(\tilde\sigma, \varphi_i, B, 
D\right)\E^{\I\int\D^4x\left[-\I {\rm tr}{\,} \ln S_{\chi}^{-1}-
\frac{1}{\tilde G}\bar B B\right]}\times$$

\begin{equation}
\E^{\I\int\int\D^4x\D^4y
\left[D^\dag (x)\left(\Delta^{-1}-\bar B S_\chi B\right)_{(x,y)}
D(y)\right]},
\end{equation}

$$
{\cal Z}={\cal N}_2\int {\cal D}\mu\left(\tilde\sigma, \varphi_i, 
B\right)\exp\left\{\I\int\D^4x\left[-\I {\rm tr}{\,} \ln 
S_\chi^{-1}-
\frac{1}{\tilde G}\bar B B+\right.\right.$$

\begin{equation}
\left.\left.\I\ln\left(1-\bar B S_\chi\Delta B\right)_
{(x,x)}\right]\right\},
\end{equation}
where the trace tr runs over Dirac, isospin, and colour indices, 
${\cal D}\mu$ denotes the integration measure of fields, $\Delta^
{-1}=-\Box-M_D^2$ is the inverse diquark propagator, and $S_\chi$ is the 
quark propagator defined by (Cf. (28) of {\em I})

\begin{equation}
S_\chi^{-1}=\I\hat D-m-\tilde\sigma-\hat{\cal A}\gamma_5.
\end{equation}
Expanding now the logarithms in power series at the one-loop level 
(see Fig.1) and performing a low-momentum expansion of Feynman 
diagrams (corresponding to a derivative expansion in configuration space), 
one describes both the generation of kinetic and mass terms of the 
composite baryon field $B$. This yields the expression 

\begin{equation}
\int\D^4x\D^4y\bar B(x)\left[-\left(\frac{1}{\tilde G}+Z_1^{-1}\frac
{\vec\tau}{2}\vec{\hat{\cal V}}+g_A\frac{\vec\tau}{2}\vec{\hat {\cal A}}
\gamma_5\right)\delta^4\left(x-y\right)-\Sigma\left(x-y\right)\right]
B(y).
\end{equation}

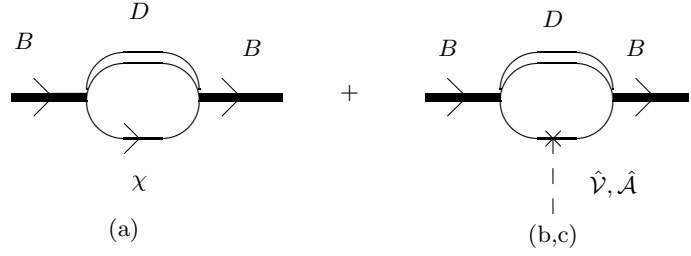
\begin{figure}[hbt]
\def\emline#1#2#3#4#5#6{%
       \put(#1,#2){\special{em:moveto}}%
       \put(#4,#5){\special{em:lineto}}}
\def\newpic#1{}

\unitlength=1mm
\special{em:linewidth 0.4pt}
\linethickness{0.4pt}
\begin{picture}(105.00,36.00)
\emline{20.00}{25.00}{1}{25.00}{25.00}{2}
\emline{20.00}{24.00}{3}{25.00}{24.00}{4}
\put(20.00,24.00){\rule{5.00\unitlength}{1.00\unitlength}}
\put(15.00,24.00){\rule{10.00\unitlength}{1.00\unitlength}}
\put(32.50,24.00){\oval(15.00,10.00)[]}
\put(32.50,25.50){\oval(15.00,10.00)[t]}
\put(40.00,24.00){\rule{11.00\unitlength}{1.00\unitlength}}
\emline{20.00}{25.00}{5}{18.00}{27.00}{6}
\emline{20.00}{24.00}{7}{18.00}{22.00}{8}
\emline{45.00}{25.00}{9}{43.00}{27.00}{10}
\emline{45.00}{24.00}{11}{43.00}{22.00}{12}
\put(32.00,36.00){\makebox(0,0)[cc]{$D$}}
\emline{32.00}{19.00}{13}{30.00}{21.00}{14}
\emline{32.00}{19.00}{15}{30.00}{17.00}{16}
\put(32.00,13.00){\makebox(0,0)[cc]{$\chi$}}
\put(18.00,32.00){\makebox(0,0)[rc]{$B$}}
\put(47.00,31.00){\makebox(0,0)[cc]{$B$}}
\put(32.00,6.00){\makebox(0,0)[rb]{(a)}}
\put(60.00,25.00){\makebox(0,0)[cc]{+}}
\put(70.00,24.00){\rule{10.00\unitlength}{1.00\unitlength}}
\put(87.50,24.00){\oval(15.00,10.00)[]}
\put(87.50,25.50){\oval(15.00,10.00)[t]}
\emline{86.00}{18.00}{17}{88.00}{20.00}{18}
\emline{88.00}{18.00}{19}{86.00}{20.00}{20}
\emline{87.00}{19.00}{21}{87.00}{17.00}{22}
\emline{87.00}{15.00}{23}{87.00}{13.00}{24}
\emline{87.00}{11.00}{25}{87.00}{9.00}{26}
\put(95.00,24.00){\rule{10.00\unitlength}{1.00\unitlength}}
\emline{75.00}{25.00}{27}{73.00}{27.00}{28}
\emline{75.00}{24.00}{29}{73.00}{22.00}{30}
\emline{100.00}{25.00}{31}{98.00}{27.00}{32}
\emline{100.00}{24.00}{33}{98.00}{22.00}{34}
\put(73.00,31.00){\makebox(0,0)[cc]{$B$}}
\put(98.00,31.00){\makebox(0,0)[cc]{$B$}}
\put(87.00,35.00){\makebox(0,0)[cc]{$D$}}
\put(95.00,13.00){\makebox(0,0)[cc]{$\hat {\cal V}, \hat {\cal A}$}}
\put(87.00,6.00){\makebox(0,0)[cc]{(b,c)}}
\end{picture}
\caption{Baryon self-energy diagram $\Sigma$ (a) and vertex diagrams (b,c) 
arising from the loop expansion of the logarithms in Eq. (4).}
\end{figure}

The vertex renormalization constant $Z_1^{-1}$ and the axial coupling 
$g_A$ arise from the low-momentum (low derivative) expansion of the 
vertex diagrams of Fig.1 b) and c), respectively. The nucleon 
self-energy $\Sigma$ has in momentum space the decomposition 
$\Sigma (p)=\hat p\Sigma_{\rm V}\left(p^2\right)+\Sigma_{\rm S}
\left(p^2\right)$. 
Its low momentum expansion generates a kinetic term $\left(\sim Z^{-1}\hat p
\right)$ and, together with the constant term $-1/\tilde G$, a mass 
$M_B$ given by the equation 

\begin{equation}
\frac{1}{\tilde G}+M_B\Sigma_{\rm V}\left(M_B^2\right)+\Sigma_{\rm S}\left(
M_B^2\right)=0.
\end{equation}
In terms of renormalized fields defined by $B=Z^{\frac{1}{2}}B_{\rm r}$, 
where $Z$ is the wave function renormalization constant satisfying the 
Ward identity 
$Z=Z_1$, we obtain from (6) the effective chiral meson-baryon 
Lagrangian [10]

\begin{equation}
{\cal L}_{\rm eff.}^{\rm MB}=\bar B_{\rm r}\left(\I\hat D-M_B\right)
B_{\rm r}-
g_A^{\rm r}\bar B_{\rm r}\gamma_\mu\gamma_5\frac{\tau_i}{2} B_{\rm r} 
{\cal A}_i^\mu,
\end{equation}
with $\hat D=\hat\partial+\I\hat{\cal V}$ and $g_A^{\rm r}=Zg_A$ 
being the renormalized axial coupling constant. 
Note that expression (8) completely coincides in structure with the 
famous phenomenological chiral Lagrangians introduced at the end 
of the Sixties when considering nonlinear realizations of chiral 
symmetry [6,7]. However, in our case these Lagrangians are not 
obtained on the basis of symmetry arguments alone, but derived from 
an underlying microscopic quark-diquark picture which allows us to 
estimate masses and coupling constants of composite hadrons. Note that,  
due to ${\cal A}_\mu^i=\frac{1}{F_\pi}\partial_\mu\varphi^i+\cdots$, the 
second term in (8) leads to a derivative coupling of the pion field 
$\vec\varphi$ with the axial-vector baryon current. In order to get rid 
of the derivative and to reproduce the standard $\gamma_5$ coupling, 
it is convenient to redefine the baryon field $B_{\rm r}\to\tilde B$ by 

\begin{equation}
B_{\rm r}=\E^{-\I g_A^{\rm r}\gamma_5\frac{\vec\tau\cdot\vec\varphi}{2F_\pi}}
\tilde B.
\end{equation}

Inserting (9) into (8) and performing a power series expansion 
in $\vec\varphi$ leads to the expression  

\begin{equation}
{\cal L}_{\rm eff.}^{{\rm MB}}=\bar{\tilde B}\left(\I\hat\partial-
M_B\right)\tilde B+g_A^{\rm r}\frac{M_B}{F_\pi}\bar {\tilde B}\I\gamma_5
\vec\tau\cdot\vec\varphi \tilde B +O\left(\vec\varphi^2\right).
\end{equation}
Obviously, we have to identify the factor in front of the interaction 
term as the pion-nucleon coupling constant $g_{BB\varphi}$,

\begin{equation}
g_{BB\varphi}=g_A^{\rm r}\frac{M_B}{F_\pi},
\end{equation}
which is nothing else than the Goldberger-Treiman relation of the 
composite nucleon. Finally, by combining the Lagrangian (8) with the 
effective chiral meson Lagrangian of the nonlinear $\sigma$ model 
(Cf. (31) of {\em I}), we obtain the complete meson-baryon Lagrangian 

\begin{equation}
{\cal L}_{{\rm eff., tot.}}^{{\rm MB}}={\cal L}_{{\rm nlin.}}^\sigma +
{\cal L}_{{\rm eff.}}^{{\rm MB}}.
\end{equation}

It is further possible to estimate magnetic moments as well as electric 
and magnetic radii of composite protons and neutrons by introducing 
electromagnetic interactions into the toy Lagrangian (1). The obtained 
pattern of predicted low-energy characteristica of nucleons has been 
shown to describe data at best qualitatively [11]. Obviously, in order 
to get better agreement with data, it is necessary to consider more 
realistic, but also more complicated models with nonlocal quark-diquark 
interactions containing both scalar and axial-vector diquarks [3-5,12]. 

\section{Further Extensions}
\subsection{Heavy Baryons with Scalar and Axial-Vector Diquarks}
The above considerations can be easily generalized to heavy-light 
baryons $B\sim \left(Qqq\right)$, where $Q=c,b$ is a heavy quark 
and the light quarks $q=u,d,s$ form scalar $(D)$ and axial-vector 
diquarks $\left(F_\mu\right)$ of the flavour group $SU(3)_{\rm F}$. The 
quantum numbers of the light diquarks with respect to the spin, 
flavour, and 
colour follow here from the decomposition 

$${\rm spin}: \frac12\times\frac12=0_{\rm a}+1_{\rm s}$$

$${\rm flavour}: 3_{\rm F}\times 3_{\rm F}=\bar 3_{{\rm F,a}}+
6_{{\rm F,s}}$$
  
$${\rm colour}: 3_{\rm c}\times 3_{\rm c}=\bar 3_{{\rm c,a}}
\left(+6_{{\rm c,s}}\right),$$
where the indices s,a refer to symmetry, antisymmetry of the respective 
wave functions under interchange of quark indices. According to the Pauli 
principle the fields of the scalar and axial-vector diquarks must then 
form a flavour (anti)triplet or sextet, respectively: $D_{\bar 3_{\rm F}}, 
F_{6_{\rm F}}^\mu$. The baryons as bound states of diquarks with a 
quark $Q$ are evidently colourless, since the product representation 
$\bar 3_{\rm c}\times 3_{\rm c}$ contains the colour 
singlet $\bbbone$. In Ref. [8] 
we have studied an extended quark-diquark model given by the Lagrangian 

\begin{equation}
{\cal L}={\cal L}_0+{\cal L}_{\rm int.},
\end{equation} 

$$
{\cal L}_0=\bar q\left(\I\hat\partial-m_0\right)q+\bar Q_{\rm v}
\I v\cdot \partial Q_{\rm v}+{\rm tr}\left[\partial_\mu D^\dag
\partial^\mu D-M_D^2 D^\dag D\right]$$

\begin{equation}
-\frac14 {\rm tr}{\,} F_{\mu\nu}^\dag 
F^{\mu\nu}+ {\rm tr}{\,} M_{\rm F}^2 F_\mu^\dag F^\mu,
\end{equation}

\begin{equation}
{\cal L}_{\rm int.}=\tilde G_1 {\rm tr}\left(\bar Q_v D^\dag\right)
\left(DQ_v\right)-\tilde G_2 {\rm tr}\left(\bar Q_v F^{\dag\mu}\right)
P_{\mu\nu}^\perp \left(F^\nu Q_v\right),
\end{equation}
where $D^{ij}, F_\mu^{ij}$ are antisymmetric/symmetric $3\times 3$ flavour 
matrices, $Q_v$ is a heavy quark spinor of 4-velocity $v_\mu$ (using 
the notation of Heavy Quark Effective Theory [13]), $P_{\mu\nu}^\perp=
g_{\mu\nu}-v_\mu v_\nu$ is a transverse projector, and $\tilde G_1, 
\tilde G_2$ are coupling constants.

In order to bilinearize the interaction term (15) one now needs two 
types of baryon fields 

$$T_v\left({\frac12}^{+}\right)\sim Q_v D,~~ S_{v\mu}\left(
{\frac12}^{+},{\frac32}^{+}\right)\sim Q_v F_\mu,$$
where $S_{v\mu}$ is a superfield of spin-symmetry partners $B_v\left(
{\frac12}^{+}\right), B_{v\mu}^*\left({\frac32}^{+}\right)$ admitting 
the decomposition 

\begin{equation}
S_{v\mu}=\frac{1}{\sqrt{3}}\gamma_5\left(\gamma_\mu-v_\mu\right)
B_v+B_{v\mu}^*.
\end{equation}
The interaction term (15) can then be rewritten as a sum of terms 
bilinear in the baryon field and a Yukawa interaction term. Integrating 
successively over the quark fields $q, Q$ and then over the diquark 
fields $D, F_\mu$ leads to determinants containing the baryon fields 
$T$ and $S$. Finally, by employing again a loop expansion and taking 
into account only lowest order derivative terms, leads to the free 
effective baryon Lagrangian of heavy flavour type 

\begin{equation}
{\cal L}_{\rm eff.}^0 ={\rm tr} \bar T_v\left(\I v\cdot \partial-
\Delta M_{\rm T}\right)T_v- {\rm tr} \bar S_{v\mu}\left(\I v\cdot
\partial-\Delta M_{\rm S}\right)S_v^\mu,
\end{equation}
where the mass differences $\Delta M_{\rm T,S}\equiv M_{\rm T,S}-m_{\rm Q}$ 
are calculable. Moreover, taking into account vertex diagrams analogously 
to those shown in Fig.1 (b,c) leads to the inclusion of interactions 
with the $SU(3)_{\rm F}$-octet of light pseudoscalar mesons $\varphi_i$ [8].

\subsection{Composite Diquarks}
For simplicity, we have considered up to now only models containing 
diquarks as elementary fields. Clearly, it is desirable to treat 
diquarks on the same footing as composite mesons as composite 
particles. This has been done in Ref. [9] by considering an extended 
NJL model for light $(q=u,d,s)$ and heavy quarks $(Q=c,b)$ containing 
2-body interactions of diquark-type $(qq)$ and $(qQ)$ 

\begin{equation}
{\cal L}_{\rm NJL}^{\rm int.}=G_1\left(\bar q^{\rm c}\Gamma^\alpha q\right)
\left(\bar q\Gamma^\alpha q^{\rm c}\right)+G_2\left(\bar q^{\rm c}
\Gamma_v^\alpha 
Q_v\right)\left(\bar Q_v\Gamma_v^\alpha q\right).
\end{equation}
Here $q^{\rm c}=C\bar q^{\cal T}$ denotes a charged-conjugated quark field, 
and $\Gamma^\alpha, \Gamma_v^\alpha$ are flavour and Dirac (spin) 
matrices. The interaction term (18) can again be rewritten in terms 
of a Yukawa coupling of light composite scalar and axial-vector 
diquark fields $D\left(0^+, 1^+\right)$ or heavy diquarks $D_v 
\left(J^{\rm P}=0^+, 1^+\right)$ to quarks. Notice that these models 
lead to a nonlocal quark-diquark interaction mediated by quark 
exchange shown in Fig.2.

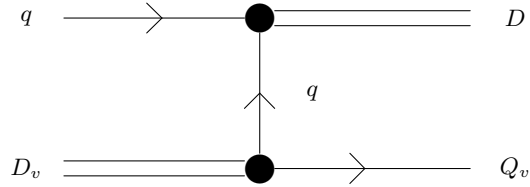
\begin{figure}[hbt]
\def\emline#1#2#3#4#5#6{%
       \put(#1,#2){\special{em:moveto}}%
       \put(#4,#5){\special{em:lineto}}}
\def\newpic#1{}

\unitlength=1mm
\special{em:linewidth 0.4pt}
\linethickness{0.4pt}
\begin{picture}(80.00,43.00)
\emline{20.00}{20.00}{1}{44.00}{20.00}{2}
\emline{20.00}{18.00}{3}{44.00}{18.00}{4}
\put(46.00,19.00){\circle*{4.00}}
\emline{48.00}{19.00}{5}{74.00}{19.00}{6}
\emline{46.00}{21.00}{7}{46.00}{37.00}{8}
\emline{46.00}{29.00}{9}{44.00}{27.00}{10}
\emline{46.00}{29.00}{11}{48.00}{27.00}{12}
\put(46.00,39.00){\circle*{4.00}}
\emline{48.00}{38.00}{13}{74.00}{38.00}{14}
\emline{48.00}{40.00}{15}{74.00}{40.00}{16}
\emline{44.00}{39.00}{17}{20.00}{39.00}{18}
\emline{33.00}{39.00}{19}{31.00}{41.00}{20}
\emline{33.00}{39.00}{21}{31.00}{37.00}{22}
\emline{60.00}{19.00}{23}{58.00}{21.00}{24}
\emline{60.00}{19.00}{25}{58.00}{17.00}{26}
\put(15.00,19.00){\makebox(0,0)[cc]{$D_v$}}
\put(80.00,19.00){\makebox(0,0)[cc]{$Q_v$}}
\put(53.00,29.00){\makebox(0,0)[cc]{$q$}}
\put(15.00,39.00){\makebox(0,0)[cc]{$q$}}
\put(80.00,39.00){\makebox(0,0)[cc]{$D$}}
\end{picture}
\caption{Quark-exchange diagram leading to a nonlocal quark-diquark 
interaction with composite diquarks.}
\end{figure}

As shown in a series of papers [3-5,9], the baryon spectrum can then 
be found by solving Faddeev-type equations for quark-diquark bound 
states. Finally, the above method of path integral hadronization 
can also be applied to investigate 3-body interactions of the type 
${\cal L}\sim \left(\bar q \bar q \bar q\right)\left(qqq\right)$ [5]. 

I hope that these few examples are sufficient to show you that 
path integral hadronization is indeed a powerful nonperturbative 
method in particle physics.

\section*{References}

\noindent
1.~Ebert, D. and Pervushin, V.N.: Teor. Mat. Fiz. {\bf 36} (1978), 759; 
Ebert, D. and Kaschluhn, L.: Nucl. Phys. {\bf B355} (1991), 123.\\
2.~Cahill, R.T., Praschifka, J., and Burden, C.J.: Aust. J. Phys. 
{\bf 42} (1989), 147, 161; Kahana, D. and Vogl, U.: Phys. Lett. 
{\bf B244} (1990), 10; Ebert, D., Kaschluhn, L., and Kastelewicz, G.: 
Phys. Lett. {\bf B264} (1991), 420.\\
3.~Cahill, R.T.: Aust. J. Phys. {\bf 42} (1989), 171.\\
4.~Reinhardt, H.: Phys. Lett. {\bf B244} (1990), 316.\\
5.~Ebert, D. and Kaschluhn, L.: Phys. Lett. {\bf B297} (1992), 367.\\
6.~Coleman, S., Wess, I., and Zumino, B.: Phys. Rev. {\bf 177} (1969), 
2239; ibid. 2247.\\ 
7.~Ebert, D. and Volkov, M.K.: Fortschr. Phys. {\bf 29} (1981), 35.\\
8.~Ebert, D., Feldmann, T., Kettner, C., and Reinhardt, H.: Z. Phys. 
{\bf C71} (1996) 329.\\
9.~Ebert, D., Feldmann, T., Kettner, C., and Reinhardt, H.: Preprint 
DESY-96-010, to appear in Int. J. Mod. Phys. {\bf A}.\\
10.Ebert, D. and Jurke, T.: Preprint HUB-EP-97/74; hep-ph/9710390 (1997).\\
11.Jurke, T.: Diploma Thesis, Humboldt University, Berlin, (1997).\\
12.Keiner, V.: Z. Phys. {\bf A354} (1996), 87.\\
13.Isgur, N. and Wise, M.: Phys. Rev. {\bf D41} (1990), 151; Neubert, M.: 
Phys. Rept. {\bf 245} (1994), 259.

\end{document}